\documentclass[submission,copyright,creativecommons]{eptcs}
\providecommand{\event}{QPL 2015} 
\usepackage{breakurl}             
\usepackage{amssymb,amsmath,amsthm,tikz,mathtools,hyperref,syntax,stmaryrd}
\theoremstyle{plain}
\newtheorem{theorem}{Theorem}
\newtheorem{lemma}[theorem]{Lemma}

\theoremstyle{definition}
\newtheorem{definition}[equation]{Definition}
\usetikzlibrary{shapes,arrows}
\tikzstyle{every loop}=[]
\newcommand{\be}{\begin{equation}}
\newcommand{\ee}{\end{equation}}
\newcommand{\bal}{\begin{align}}
\newcommand{\eal}{\end{align}}
\newcommand{\lsem}{\llbracket}
\newcommand{\rsem}{\rrbracket}
\newcommand{\DD}{\mathcal{D}}
\newcommand{\LL}{\mathcal{L}}
\newcommand{\State}{\mathsf{State}}
\newcommand{\rarr}{\rightarrow}

\title{\textsc{DEMONIC} programming: a computational language for single-particle equilibrium thermodynamics, and its formal semantics}
\author{
Samson Abramsky \hspace{2cm} Dominic Horsman
\institute{Department of Computer Science, University of Oxford, Parks Road, Oxford, OX1 3QD, UK}
\email{\{samson.abramsky,clare.horsman\}@cs.ox.ac.uk}
          }  
\def\titlerunning{\textsc{demonic} for single-particle equilibrium thermodynamics}
\def\authorrunning{
S. Abramsky \& D. C. Horsman
}
\begin{document}
\maketitle

\begin{abstract}
Maxwell's Demon, `a being whose faculties are so sharpened that he can follow every molecule in its course', has been the centre of much debate about its abilities to violate the second law of thermodynamics. Landauer's hypothesis, that the Demon must erase its memory and incur a thermodynamic cost, has become the standard response to Maxwell's dilemma, and its implications for the thermodynamics of computation reach into many areas of quantum and classical computing. It remains, however, still a hypothesis. Debate has often centred around simple toy models of a single particle in a box. Despite their simplicity, the ability of these systems to accurately represent thermodynamics (specifically to satisfy the second law) and whether or not they display Landauer Erasure, has been a matter of ongoing argument. The recent Norton-Ladyman controversy is one such example.

In this paper we introduce a programming language to describe these simple thermodynamic processes,  and give a formal operational semantics and program logic as a basis  for formal reasoning about thermodynamic systems. We formalise the basic single-particle operations as statements in the language, and then show that the second law must be satisfied by any composition of these basic operations. This is done by finding a computational invariant of the system. We show, furthermore, that this invariant requires an erasure cost to exist within the system, equal to $kT\ln 2$ for a bit of information: Landauer Erasure becomes a theorem of the formal system. The Norton-Ladyman controversy can therefore be resolved in a rigorous fashion, and moreover the formalism we introduce gives a set of reasoning tools for further analysis of Landauer erasure, which are provably consistent with the second law of thermodynamics.
\end{abstract}

\section{Introduction}

The thermodynamics of computation is an important area within both quantum and classical reversible computation \cite{leffrex}. It concerns the interaction between physical processes, specifically those involving \textit{entropy} change, and logical processes involving creation and processing of \textit{information} \cite{maroney05,ladyman07}. 
For quantum computing, arguments around thermodynamic consequences of logical operations impact on the ability to build functioning computers using error correction \cite{vlatko}, and on the role played by entanglement in computation \cite{lloyd}. In the classical arena, thermodynamic considerations are viewed as having important implications for the possibility of fully reversible computational processes \cite{toffoli}. In both quantum and classical cases, the \emph{second law of thermodynamics} (that overall entropy cannot decrease with time) is seen as having profound consequences for the information processing abilities of physical systems.

A key concept in the thermodynamics of computing is \textit{erasure}. An important element is the hypothesis due to Landauer that erasing 1 (qu)bit of information entails a minimum entropy cost of $k\ln 2$ where $k$ is the Boltzmann constant 
\cite{landauer}. Landauer's hypothesis was introduced to get around the problem of Maxwell's Demon: a diabolic entity who could apparently violate the second law of thermodynamics within a closed system \cite[Ch 12]{maxwell}. It has never been proven, however, that Landauer Erasure is a process that actually ocurrs in the physical world. A common line of argument is to give toy model thermodynamic systems (frequently only a few particles in a box), and to try and argue the inevitability of an entropy cost for an erasure. One such line of argument has been given recently by Ladyman et al \cite{ladyman07,ladyman13}, and fiercely criticised by Norton \cite{norton11}. They both use a `library set' of basic configurations and allowed transitions, and then argue whether certain composite transitions are allowed from this basic set. In particular, there is disagreement over whether the basic transitions, with or without a Maxwell Demon type entity, themselves allow violations of the second law. This is a crucial question, as the logic of the arguments which use the basic transition systems is as follows. First add a Demon to the library set, and show that the second law is violated. Then modify the Demon to include the erasure cost specified by Landauer Erasure, and from there recover the second law. If the basic transitions themselves violate the second law, then they are not only a bad model for thermodynamics, they cannot logically be used to show the necessity of Landauer's erasure cost.

In this paper we use the Computer Science methods of formal semantics and verification to demonstrate that the set of basic transitions discussed by Ladyman and Norton is fully consistent with thermodynamics. Our key step is to define a programming language for single-particle equilibrium thermodynamics, termed \textsc{demonic}, giving syntax and an operational semantics for the set of allowed operations, concentrating here on single systems. Allowed transitions are given, and we show that some are of necessity nondeterministic. Sequential and parallel composition rules are given. In order to prove properties of the thermodynamic processes which can be described in our language, we use \emph{Hoare logic} \cite{hoare}. We verify that a certain property is a computational invariant of the system, and use it to demonstrate that the second law (in its Kelvin formulation) holds for every process which can be described by a program in our language. Moreover, we demonstrate that a controversial transition (the Norton `reset' procedure \textsc{NReset}) is \textit{not} derivable from the basic library set, as it violates this invariant. Adding this transition to the system allows a transition  to be constructed which violates the second law. \textsc{NReset} is therefore rejected. A further consideration of the computational invariant demonstrates that Landauer Erasure itself is also a theorem of the system: there is of necessity an entropic work cost to information erasure in this formal system. We therefore have a formal model for single-particle thermodynamics that is provably consistent with the second law, and where Landauer Erasure is a provable consequence of any erasure event.

\section{Single-particle equilibrium thermodynamics}

The second law of thermodynamics, in one of its formulations, states that within a closed system the quantity \textit{entropy} is 
non-decreasing over time. Some processes are only allowed to operate in one temporal direction. This appears to conflict with underlying reversible microdynamics, either quantum or classical Newtonian. Since the formulation of thermodynamics, toy model thermodynamic systems have been used in debates around these foundational issues. The simplest example is a single particle in a box, in contact with a heat bath. The thermodynamic variables can be easily specified, while at the same time the microdynamics for the single particle are easily comprehensible. While there are important issues around whether this system can actually be said to have a thermodynamics (usually considered a theory of large numbers of particles), it is a useful tool; and it is this form of system that has been considered by a large number of authors, starting with Szilard \cite{szilard}. By formalising this system, we also give a starting-point for a similar semantics of many-particle thermodynamic systems.

The fundamental system under consideration is a single particle gas in a box, connected to an external heat bath. The heat bath keeps the entire system at a specific temperature $T$; this is a constant and not a variable. The thermodynamic variables for the system are: pressure of the gas (i.e. the force exerted on the walls of the box by the continuous motion of the single particle) $p$, volume occupied by the gas $V$, and entropy of the system $H$. There is a removable partition in the middle of the box, and two pistons (left and right) which can either compress the gas and input work $-W$ into the system (\textit{isothermal compression}) or expand under pressure of the gas thereby extracting work $W$ from the box (\textit{isothermal expansion}).

There are two fundamental equations of state for the system:
$$ \frac{pV}{T} = \mathrm{const} \hspace{3cm} H_\mathrm{out} - H_\mathrm{in} = k \ln \frac{V_{\mathrm{out}}}{V_{\mathrm{in}}} $$


\noindent where again $k$ is the Bolzmann constant, and where `out' and `in' refer respectively to output and input values of parameters for a given transition. The first means that pressure and volume are not independent at a fixed temperature; only one variable need be specifed. The second equation means that a decrease in system volume of a half decreases the entropy of the system by $k\ln2$. A doubling of the system volume increases the entropy by the same amount. In isothermal expansion from volume $V \rightarrow 2V$, the amount of work done by the gas on the piston is $W = kT\ln2$. In isothermal compression $2V \rightarrow V$, the same amount of work is input to the gas by the piston.

The entropy $H$ of the system is defined in terms of the probability for the particle position. With a removable partition, the box is divided into two regions: left and right. This division is considered as existing even when the partition and pistons are all absent, and the particle occupies both positions with equal probability. If the particle has a probability of being in the left-hand configuration of $p_L$ and a corresponding right-hand probability $p_R = 1-p_L$ then $H = -k (p_L \ln p_L + p_R \ln p_R) $. As a consequence, if the particle has either $p_L=0,1$ then $H=0$. If $p_L = \frac{1}{2}$ then the entropy of the box is $H=k\ln 2$. Note that the units of entropy are the units of work/temperature.

The \textbf{Kelvin statement of the second law} of thermodynamics says that there exists no allowed transition that returns the system to its starting configuration, $\Delta H =0$, with a net extraction of work, $\Delta W > 0$.

The box described above has an information-theoretic representation as a two-state system: the particle can be either on the left or the right, corresponding to `0' and `1' respectively. One bit of data can therefore be encoded in its position. \textbf{Landauer's Erasure Hypothesis} conjectures that an erasure of 1 (qu)bit of information must be paid for by a corresponding increase in entropy of the system of $\Delta H = k\ln2$. Equivalently, that erasing one bit of information requires work $W = kT\ln2$ to perform.

An \textbf{erasure operation} returns the system to a \textit{known} state: erasure is not the same as making the system maximally unknown. Erasure in the box system given here is defined as the process $p_L \rightarrow p_L=1$ i.e. the system returns to the particle in the left-hand configuration (bit value 0) with probability 1. This is sometimes referred to as a `reset' operation.

Landauer and others have conjectured that only the presence of the entropy cost of Landauer Erasure prevents the second law of thermodynamics being broken in certain transitions embodied in the example of \textbf{Maxwell's Demon}. In a system of many particles, which have a distribution of velocities, Maxwell's Demon sits at a hole in the partition and observes the speed of particles that are near the hole. It permits high-velocity ones only into the left and keeps low-velocity ones to the right by opening or shutting a frictionless cover. Without doing work on the system, then, it creates a temperature gradient that can do external work before returning to the original configuration, thus violating the second law. By noting that the Demon must erase its memory to return to its original configuration, Landauer postulated that the Erasure Hypothesis preserves the second law.

The \textbf{Norton-Ladyman controversy} concerns the existence of Landauer Erasure in the box system described above, and the status of the second law in respect to its allowed transitions. The arguments are:

\noindent \textbf{Ladyman1}: Given a set of allowed configurations and transitions for the single-particle box, unless Landauer Erasure exists then the second law is violated \cite{ladyman07}.

\noindent \textbf{Norton1a}: The set of allowed transitions used in Ladyman1 themselves violate the second law, regardless of the energy cost of erasure. Therefore they cannot be used to determine the correctness or otherwise of Landauer's Hypothesis \cite{norton11}.

\noindent \textbf{Norton1b}: These allowed transitions can also be used to construct a reset operation \textsc{NReset} that erases a bit of information without an energy cost \cite{norton11}.

\noindent \textbf{Ladyman2a}: The allowed transitions do not violate the second law \cite{ladyman13}.

\noindent \textbf{Ladyman2b}: The addition of \textsc{NReset} to the allowed transitions \textit{does} permit such violations \cite{ladyman13}.

We now give a solution to this debate by formalising the basic box and transition system in the newly-defined language \textsc{demonic}. We prove that Ladyman2a and Ladyman2b are correct statements in this formal system (although for different reasons than given by Ladyman et al.). Moreover we show that Landauer's Hypothesis is a theorem of the system.

\section{Syntax and semantics for \textsc{demonic}}

In order to construct a formal model of the single-particle box system described in the previous section, we first define the syntax of the thermodynamic state variables, and then give the transition rules of the system in terms of the semantics of allowed basic operations. We will then be able to use formal verification methods to prove theorems about any possible transition within this system.

 A transition or a cycle is given as a sequence of basic operations in advance -- that is, as a program. There is no place within this model for online agents to make decisions within  the system. This is in contrast with the standard method of argument within either physics or philosophy, where a common statement includes elements such as ``if you know the particle is in the left-hand side then \ldots''. These elements can cause a great deal of confusion; by eliminating them, and requiring precise offline specification of the system's processes, our formalisation makes the logical structure of the arguments much clearer, and brings a greater degree of precision to the debate. 

A feature of this methodology, in which all basic operations must be specified for any initial configuration, without an agent deciding whether or not to apply a certain box at a certain time, is that all states must be considered as possible inputs. This particularly important  when the action of a piston is considered. This method of reasoning about thermodynamic systems uses the tools of  programming language theory in order to formalise the logic of processes, in the same way that predicate calculus formalises the logic of statements.
 
To formalise the thermodynamic system of the previous section, we begin by isolating the necessary variables.  A full specification of the thermodynamics state will be given by the volume of the gas $V$, the position of the particle $x$, and entropy $H$. We will also wish to keep track of the total amount of work extracted from the system, which we will label $W= \mbox{work out} - \mbox{work in}$. The basic transitions are:

\begin{itemize}
\item \textbf{PartIn}: A partition is inserted into the centre of the box, thereby halving the volume of the gas.
\item \textbf{PartOut}: The partition is removed. This doubles the volume of the gas.
\item \textbf{LPistIn}: A piston is inserted on the left as far as the centre of the box. If a partition is present then the volume of the gas does not change. If a partition is absent then the volume of the gas halves and the the piston inputs work $kT\ln2$ to the system (isothermal compression). Note that a gas cannot be compressed to zero volume (this is unphysical as it would entail infinite pressure on the piston and therefore infinite energy to perfom).
\item \textbf{RPistIn}: As LPistIn, only the piston is inserted on the right.
\item \textbf{LPistOut}: A piston is removed to the left. If a partition is present then the volume of the gas does not change. If a partition is absent then the volume of the gas doubles, and the piston extracts work $kT\ln2$ from the system (isothermal expansion)
\item \textbf{RPistOut}: As RPistIn, only the piston is removed on the right.
\end{itemize}

We formalise the state space of the system as a cartesian product, with typical \emph{state variable} $s$: $ s = (X,A,I,w) \in \mathbb{T} \times \mathbb{B} \times \mathbb{B} \times \mathbb{Z} $. The field $X$ is defined as the probability for the particle to be in the left-hand-side of the box, $p_L$. We consider in this present work the simplest possible set of probabilities for the particle location: either definitely left or right (1 or 0), or else a uniform distribution between the two. Therefore $X \in \mathbb{T} := \{ 0,\frac{1}{2},1\}$. The variables $A,I$ report the presence or absence of a partition and a piston respectively. $W = w kT\ln2$ is as above, the total work extracted from the system; the field $w$ is an integer owing to the operations we will be allowing later.  These values permit the volume of the gas to be deduced, and also the entropy. An assignment statement for each field given the state is written eg. $s.X := \frac{1}{2}$, which assigns the value $\frac{1}{2}$ to the $X$ field in state $s$.

A basic \textit{statement} in the language is either an assignment statement or a $\mathtt{skip}$. There are three methods of combining statements: sequential composition, if-then-else statements, and probabilistic choice, written as $S_1 \oplus S_2$. 

\begin{table}
\hspace{4cm}\parbox{11cm}{
LProb $ \ ::= \mathbb{T} :=\  \mathtt{0} \ | \ \mathtt{1/2} \ | \ \mathtt{1} $\\
Part $ \ ::= \ \mathbb{B} := \ \mathtt{true} \ | \ \mathtt{false} $ \\
Pist $ \ ::= \ \mathbb{B} $\\
WUnit $ \ ::= \ \mathbb{Z} $\\
Field $ \ :: = \ LProb \ | \ Part \ | \ Pist \ | \ WUnit $\\
Fieldname $ \ ::= \ \mathtt{X} \ | \ \mathtt{A} \ | \ \mathtt{I} \ | \ \mathtt{w} $ \ \ (where $W=wkT\ln2$)\\
$s \in$ State $ \ ::= \ (LProb, Part, Pist, WUnit)$\\
BExp $ \ ::= \ \mathbb{B} \ | \ State.A \ | \ State.I$\\
$S \in$ Statement $ \ ::= \ S_1 ; S_2 \ | \ \ S_1 \oplus S_2 \ | \ State.Fieldname := Field  \\
{}\hspace{3cm} | \ \mathtt{if} \ BExp \ \mathtt{then} \ S_1 \ \mathtt{else} \ S_2 \ | \ \mathtt{skip}$
}
\caption{\textsc{demonic} syntax.}\label{demsyn}
\end{table}

\begin{table}{
\begin{flalign}
\mathrm{(assign)} \hspace{1cm} &\langle x := a, \ s\rangle \Rightarrow \langle\mathtt{skip}, s[x\mapsto a]\rangle\nonumber\\
\mathrm{(comp1)} \hspace{1cm}  &\frac{\langle S_1 , \ s \rangle \Rightarrow_p \langle S'_1, s'\rangle}{\langle S_1;S_2, \ s \rangle \Rightarrow_p \langle S'_1 ;S_2 , \ s' \rangle}\nonumber\\
\mathrm{(comp2)} \hspace{1cm}  &\langle \mathtt{skip};S, \ s \rangle \Rightarrow \langle S, \ s \rangle\nonumber\\
\mathrm{(if1)}  \hspace{1cm}  & \langle \mathtt{if} \ B  \ \mathtt{then} \  S_1 \ \mathtt{else} \ S_2, \ s \rangle \Rightarrow \langle S_1,\ s\rangle \  \mathrm{if} \ \lsem B \rsem s =\mathtt{true}\nonumber\\
\mathrm{(if2)}  \hspace{1cm}  & \langle \mathtt{if} \ B  \ \mathtt{then} \  S_1 \ \mathtt{else} \ S_2, \ s \rangle \Rightarrow \langle S_2,\ s\rangle \  \mathrm{if} \ \lsem B \rsem s =\mathtt{false}\nonumber\\
\mathrm{(prob1)} \hspace{1cm} &\langle S_1 \oplus S_2, \ s \rangle \Rightarrow_{1/2} \langle S_1, \ s \rangle  \nonumber\\
\mathrm{(prob2)} \hspace{1cm} &\langle S_1 \oplus S_2, \ s \rangle \Rightarrow_{1/2} \langle S_2, \ s \rangle  \nonumber
\end{flalign}\vspace{-1cm}}
\caption{Operational semantics for \textsc{demonic}.}\label{demsem}\end{table}

The full syntax is given in table \ref{demsyn}. The operational semantics is given in table \ref{demsem}.
The semantics defines a probabilistic transition system on \emph{configurations}, which are pairs $\langle S, s\rangle$, where $S$ is a statement and $s$ a state, using the standard methods of structural operational semantics \cite{plotkin1981structural}.
Transitions have the form $\langle S, s\rangle \Rightarrow_p \langle S', s'\rangle$, where $p \in [0, 1]$ is a probability. The \emph{final configurations}, those of the form $\langle \mathtt{skip}, s\rangle$, have no outgoing transitions; for all other configurations, the probabilities of the outgoing transitions sum to 1.
We use the convention that if the probability label is omitted, the transition occurs with probability 1.

We shall assume a standard evaluation function, which given a boolean expression $B$ and a state $s$, returns a boolean value $\lsem B \rsem s$. We also use the standard update function on states, $s[x \mapsto a]$, which given a state $s$, a component $x$ and a value of the appropriate type $a$, updates the $x$ component of $s$ with $a$, leaving the other components unchanged.
For background on these notations, see e.g.~\cite{hennessy1990semantics}.

 This is the basic structure of \textsc{demonic}. The language itself does not encode any information about allowed transitions, or any kind of thermodynamic physical laws. We now turn to how these are defined.
 
\section{Thermodynamic operations} 
 
In order to formalise the dynamics and rules of thermodynamics, we now define the basic allowed transitions, given above, as statements in \textsc{demonic}.

 \textbf{PartIn} $\ =_{def}\ (s.A := \mathtt{true})$\\
Inserting a partition changes the partition flag variable $A$ to `true', but changes no other state variable.

\textbf{PartOut} $\ =_{def}\ $ if $(s.A = \mathtt{true})$ \\
\indent \hspace{3cm} then (if $(s.I = \mathtt{false})$\\
\indent \hspace{4.5cm} then $(s.X := \tfrac{1}{2}) \ $ and $ (s.A := \mathtt{false}) \ $\\
\indent \hspace{4.5cm} else $(s.A := \mathtt{false}) $ )\\
\indent \hspace{3cm} else $\mathtt{skip}$\\
Removing a partition does nothing unless the partition is present. In that case, if there is not also a piston, the particle now has the entire box to be found in (so $p_l=X=0.5$).

\textbf{LPistOut} $\ =_{def}\ $ if $(s.I = \mathtt{false})$ or $\neg (s.X = 0)$ \\
\indent \hspace{3cm} then $\mathtt{skip}$\\
\indent \hspace{3cm} else ( if $(s.A = \mathtt{true}) \ $ \\
\indent \hspace{4.5cm} then $(s.I := \mathtt{false}) \ $ \\
\indent \hspace{4.5cm} else $(s.I := \mathtt{false}) \ $ and $(s.X := \frac{1}{2}) \ $ and $(s.w := w+1)$ )\\
Removing a piston to the left can only be done if there is a piston, and the particle is on the right hand side. If there is also a partition then no other state variable change. If there is no partition, then isothermal expansion occurs by pressure of the particle on the piston, and a unit of work is extracted from the system as the volume doubles.

\textbf{RPistOut} $\ =_{def}\ $ if $(s.I = \mathtt{false})$ or $\neg (s.X = 1) \ $  \\
\indent \hspace{3cm} then $\mathtt{skip}\ $ \\
\indent \hspace{3cm} else ( if $(s.A = \mathtt{true}) \ $ \\
\indent \hspace{4.5cm} then $(s.I := \mathtt{false}) \ $ \\
\indent \hspace{4.5cm} else $(s.I := \mathtt{false}) \ $ and $(s.X := \frac{1}{2}) \ $ and $(s.w := w+1) \ $ )\\
Removing a piston to the right is by symmetry with removing to the left.

Inserting a piston is a more complicated operation, and requires the probabilistic composition operation $\oplus$. It is probabilistic as we are using the programming style of reasoning, so have to define the allowed operations for any possible input. In the case of a piston, this includes inserting a piston such that an attempt is being made to reduce the volume of the gas to zero (for example a case where there is a partition and $p_L =0$ and a piston is inserted on the left). This is unphysical, as it would require an infinite amount of energy to perform. We define the amount of energy that is lost in the attempt to perform this unphysical operation as $W_c = w_c kT \ln 2$. The left piston operation is given as

\textbf{LPistIn} $\ =_{def}\ $ if $(s.X = 1)\ $  \\
\indent \hspace{2.5cm} then $(s.w := w-w_c)\ $ \\
\indent \hspace{2.5cm} else ( if $(s.X = 0)\ $ \\
\indent \hspace{3.5cm} then $(s.I := \mathtt{true}) \ $ \\
\indent \hspace{3.5cm} else ( if $ (s.A = \mathtt{false})\ $ \\
\indent \hspace{4.5cm} then $(s.X := 0)\ $ and $ (s.w := w-1) \ $ and $(s.I := \mathtt{true})\ $ \\
\indent \hspace{4.5cm} else $[(s.X := 0) \ \mathrm{and} \ (s.I := \mathtt{true})] \oplus [(s.X := 1) \ \mathrm{and} \ (s.w := w-w_c)]$))

\noindent If the particle known to be on the left, then there must be a partition or other piston keeping it there, and so inserting a left piston will be an attempt to compress to zero volume, incurring the work cost $w_c$. If the particle is known to be on the right, again a piston or partition must be keeping in there, and so inserting a left piston does no work. If the particle position is unknown, if there is no partition then isothermal compression occurs, halving the volume and inputting work $kT\ln 2$ to the system. If a partition is present, half the time the particle will be on the right, with subsequent actions, and half the time on the left (incurring the cost $w_c$ for failing to inset the piston as the action would be unphysical). 

We now wish to discover exactly what amount of work the `failure cost' $W_c$ is equal to; and whether it is, in fact, nonzero. We have argued on physical grounds that it must be some non-zero amount as the piston is attempting to compress the gas to zero volume, an action that would require infinite energy to perform. As we are taking the work input of the piston to be given in integer units of $kT\ln 2$, the obvious value for $W_c$ would be one such unit, and so $w_c=1$. 

To make this argument more precise, consider \textbf{Cycle} $=_{def}$ PartIn ; LPistIn ; PartOut ; LPistOut. The action this on the input state $s_{in} = \left( \frac{1}{2}, \ \mathtt{false}, \ \mathtt{false}, \ w_0 \right)$ can be stated in words as a very simple set of operations. Start with the particle completely delocalised in the box with neither pistons nor partition, ie $X=\frac{1}{2}$. Then insert a partition (\textbf{PartIn}). The particle probability of being left or right has not changed. Then insert a piston to the left. With a probability of one half this is acting on an empty half of the box, and so no work is required. With a probability of one half, however, this is an attempt to compress the gas to zero volume, which fails: no piston is inserted, and the failure cost $W_c$ (which may be zero) incurred. Then the partition is removed. Half the time (corresponding to failure to insert the piston) the particle then returns to being fully delocalised (the original configuration). The other half of the time the removal of the partition allows the particle to act on the piston and remove it via isothermal expansion, doing work $W=kT\ln2$ on the piston before returning to the original configuration.

Therefore, the result of this cycle is that half the time work $kT\ln2$ is extracted from the system before returning it to the original configuration, and the other half of the time work $W_c$ is input to the system (the failure cost of piston insertion). Regardless of any formal statement of the second law, one can see that if, say, $W_c = 0$ then we have a perpetual energy-extraction machine. We can see in more detail what $w_c$ should be by considering the effect of $\langle \textbf{Cycle},s_{in} \rangle$ in \textsc{demonic}. This branches, as it contains a probabilistic operation. The two derivations are:
\begin{align} \langle PartIn ; LPistIn ; PartOut ; & LPistOut , \ (\tfrac{1}{2}, \ F, \ F, \ w_0) \rangle & \nonumber \\
& \Longrightarrow \ \ \langle LPistIn ; PartOut ; LPistOut, \  (\tfrac{1}{2}, \ T, \ F, \ w_0) \rangle\ \nonumber\\
& \Longrightarrow_{1/2}  \ \ \langle PartOut ; LPistOut, \  (0, \ T, \ T, \ w_0)\rangle\ \nonumber\\
& \Longrightarrow  \ \ \langle LPistOut, \  (0, \ F, \ T, \ w_0)\rangle\ \nonumber\\
& \Longrightarrow  \ \ \langle skip, \  (\tfrac{1}{2}, \ F, \ F, \ w_0+1) \rangle\ \nonumber\end{align}
\begin{align}
\langle PartIn ; LPistIn ; PartOut ; & LPistOut , \ (\tfrac{1}{2}, \ F, \ F, \ w_0) \rangle & \nonumber \\
& \Longrightarrow  \ \ \langle LPistIn ; PartOut ; LPistOut, \  (\tfrac{1}{2}, \ T, \ F, \ w_0) \rangle\ \nonumber\\
& \Longrightarrow_{1/2}  \ \ \langle PartOut ; LPistOut, \ (1, \ T, \ F, \ w_0 - w_c) \rangle\ \nonumber\\
& \Longrightarrow  \ \ \langle LPistOut, \   (\tfrac{1}{2}, \ F, \ F, \ w_0 - w_c) \rangle\ \nonumber\\
& \Longrightarrow  \ \ \langle skip, \  (\tfrac{1}{2}, \ F, \ F, \ w_0 - w_c) \rangle\ \nonumber
\end{align}

The expected amount of work extracted by the end of this cycle is the weighted sum of the two work outcomes, $W_e = (w_0 + \tfrac{1}{2}(1-w_c))kT\ln 2$. To preserve the second law we require $W_e \le W_o$, i.e. $w_c \ge 1$. We now make the choice that the set of transitions given by \textbf{Cycle} not only preserves the second law, but does not require the input of work. Given this choice, $w_c = 1$. The full statements for \textbf{LPistIn} and \textbf{RPistIn} are then

\textbf{LPistIn} $\ =_{def}\ $ if $(s.X = 1)\ $  \\
\indent \hspace{2.5cm} then $(s.w := w-1)\ $ \\
\indent \hspace{2.5cm} else ( if $(s.X = 0)\ $ \\
\indent \hspace{3.5cm} then $(s.I := \mathtt{true}) \ $ \\
\indent \hspace{3.5cm} else ( if $ (s.A = \mathtt{false})\ $ \\
\indent \hspace{4.5cm} then $(s.X := 0)\ $ and $ (s.w := w-1) \ $ and $(s.I := \mathtt{true})\ $ \\
\indent \hspace{4.5cm} else $[(s.X := 0) \ \mathrm{and} \ (s.I := \mathtt{true})] \oplus [(s.X := 1) \ \mathrm{and} \ (s.w := w-1)]$))

\textbf{RPistIn} $\ =_{def}\ $ if $(s.X = 0)\ $  \\
\indent \hspace{2.5cm} then $(s.w := w-1)\ $ \\
\indent \hspace{2.5cm} else ( if $(s.X = 1)\ $ \\
\indent \hspace{3.5cm} then $(s.I := \mathtt{true}) \ $ \\
\indent \hspace{3.5cm} else ( if $ (s.A = \mathtt{false})\ $ \\
\indent \hspace{4.5cm} then $(s.X := 1)\ $ and $ (s.w := w-1) \ $ and $(s.I := \mathtt{true})\ $ \\
\indent \hspace{4.5cm} else $[(s.X := 1) \ \mathrm{and} \ (s.I := \mathtt{true})] \oplus [(s.X := 0) \ \mathrm{and} \ (s.w := w-1)]$))


\section{The second law is an invariant of the system}

We have a full specification of the basic allowed operations within single-particle thermodynamics, in our computational language. The task now is to prove that any composition of these basic statements (PartOut etc.) within \textsc{demonic} will satisfy the second law; that is, that the basic operations can be used to reason about violations of the second law without themselves violating it.

We shall now provide a general setting for proving that certain properties are invariants of a class of thermodynamic processes, which we will then go on to apply to the second law. Firstly, we write $\DD(\State)$ for the set of probability distributions of finite support on the set $\State$ of states.
A statement $S$ induces a map $\tau_S : \State \rarr \DD(\State)$:
\[ \tau_S(s)(s') \; := \; \sum \{ p \mid \exists S'. \, \langle S, s \rangle \Rightarrow_p \langle S', s' \rangle \} \]
This lifts to a map $\DD(\State) \rarr \DD(\State)$ using the Kleisli extension of the distribution monad \cite{giry1982categorical}. We also write $\tau_S$ for this lifted map.

Now let $P$ be a predicate $P \subseteq \DD(\State)$. We say that $P$ is an \emph{$S$-invariant} if for all distributions $d \in \DD(\State)$:
\[ d \in P \; \Rightarrow \; \tau_S(d) \in P . \]
Let $\LL$ be a set of statements defining a class of thermodynamic processes. We say that $P$ is an $\LL$-invariant if it is an $S$-invariant for every $S$ in $\LL$. We shall be interested in proving $\LL$-invariance where $\LL$ is the closure under sequential composition of the set of basic thermodynamic processes defined in the previous section.

As an example of the kind of predicate we may consider, we have $P(d) \; \equiv \; \langle X \rangle \leq 0$, where $\langle X \rangle \; := \; \sum_{s \in \State} d(s)s.X$,
the expected value of the $X$-component of the state in the distribution $d$.

A \textit{computational invariant} for the basic transitions is therefore a statement that is true after each transition if it is true before. Note that a computational invariant is automatically also preserved through composition. Using the definition for entropy of a variable $x$, $H(x) = -kT(x\ln x + (1-x)\ln (1-x))$, we show in the Appendix that the following is a (probabilistic) computational invariant for the basic transitions defined above:
\begin{equation} \langle wk\ln 2 \rangle - \frac{1}{2} \left( \langle H(X) \rangle + H(\langle X \rangle ) \right) \le 0 \label{inc} \end{equation}

\noindent Every sequence of basic transitions will satisfy this inequality at every step if it satisfies it at the start.

Written for the \textsc{demonic} system, the second law becomes a statement disallowing certain sequences of transitions $\gamma$:

\textbf{Kelvin statement of the second law:} $\ \not \exists \gamma : \ (X_0, \ A_0, \ I_0, \ w_0) \xrightarrow{\gamma} (X_0, \ A_0, \ I_0, \ \langle w_f \rangle > w_0) $.

\noindent The variable $w$ counts the work extracted from the system, and is transparent to the addition or subtraction of a constant. The initial counter for the work extracted can therefore be set to an arbitrary number, as long as the rest of the cycle counts from that number. 
We define the zero-point of the work counter to be
\begin{equation} w_0  = \frac{1}{2k\ln 2} \left( \langle H(X_0) \rangle + H(\langle X_0 \rangle ) \right) \end{equation}

\noindent This satisfies the invariant statement with an equality. After any sequence of transitions that return the system to the original configuration, the invariant states that $ \langle w_f \rangle k\ln 2 - \frac{1}{2} \left( \langle H(X_0) \rangle + H(\langle X_0 \rangle ) \right) \le 0 $, which straightforwardly implies $ \langle w_f \rangle \le w_0 $. The second law in its Kelvin formulation is then a theorem of any system whose statements satisfy the computational  invariant \eqref{inc}. The basic transitions (PartIn, PartOut, LPistIn, LPistOut, RPistIn, RPistOut) satisfy the invariant; therefore any sequence of these operations cannot violate the second law.

\section{The Ladyman \textsc{Shift} and the Norton \textsc{NReset}}

The set of allowed transitions given above is the core of the generally-agreed operations for single-particle thermodynamics. We now turn to two other operations that have been under debate in the Norton-Ladyman controversy: the \textsc{Shift} operation of Ladyman et. al., and Norton's \textsc{NReset}. We show that, while \textsc{Shift} satisfies the invariant \eqref{inc}, and hence does not permit violations of the second law, the reset operations \textsc{NReset} does \textit{not} satisfy the invariant, and permits transitions forbidden by the second law.

The \textsc{Shift} operation is a use of two pistons to shift a particle from the left-hand-side of the box to the right-hand-side (a similar right-to-left operation can be constructed by symmetry)\cite{ladyman13}. It is straightforward to show that this operation satisfies the invariant, as it can be constructed out of the basic transitions for the system: \textsc{Shift} $=_{def}$ RPistIn ; PartOut ; RPistOut ; LPistIn ; PartIn ; LPistOut.
If \textsc{Shift} operates on a configuration where $X_0 = 1$, the particle is shifted to the right-hand side $X=0$ with no work cost. If $X_0=0$ then it incurrs a cost $w-1$ when the RPistIn operation fails. When $X_0=\tfrac{1}{2}$, this is a reset operation, and costs $w-1$ to perform. 

The Norton reset operation, \textsc{NReset}, is a different type of reset. The basic operation is to see whether the particle is on the right-hand-side, and if it is then to use \textsc{Shift} to move it to the left with no work cost. Debate has centred around both whether this is a permitted operation, and whether it can be constructed from the other permitted operations. We can show now that it cannot be so constructed, as it violates the invariant, and that in violating the invariant it enables second-law-violating transitions to be implemented. The Norton reset operation is

\textbf{NReset} $\ =_{def}\ $ if $(s.X = 0)\ $  \\
\indent \hspace{3cm} then $(s.X := 1)\ $ and $(s.A := 1)\ $ \\
\indent \hspace{3cm} else ( if $(s.X = 1)\ $ \\
\indent \hspace{4.5cm} then \texttt{skip} \\
\indent \hspace{4.5cm} else ( if $(s.X = \tfrac{1}{2})\ $ \\
\indent \hspace{6cm} then $(s.X := 1) $ and $(s.A := 1)\ $ \ )) 

This is a good example of the utility of \textsc{demonic} in reasoning about thermodynamic systems: while the description given by Norton can appear innocuous, its formalisation clearly shows that it is a much more powerful operation than the others we have considered. 

In fact, the operation of \textsc{NReset} can be reduced to: \textbf{NReset} $\ = \ (s.X := 1) $ and  $(s.A := 1)\ $, which straightforwardly violates the invariant. Consider for example the transition $ \langle NReset, \ (\tfrac{1}{2} , F, F, 1 ) \rangle  \ \Longrightarrow \langle skip, \ (1 , T, F, 1 ) \rangle$. Before the transition we have $ \langle wk\ln 2 \rangle - \frac{1}{2} \left( \langle H(X) \rangle + H(\langle X \rangle ) \right) = k\ln2 - k\ln 2 = 0 \le 0$. After the transition the invariant is violated: $ \langle wk\ln 2 \rangle - \frac{1}{2} \left( \langle H(X) \rangle + H(\langle X \rangle ) \right) = k\ln2 - 0  > 0 $. Furthermore, there is a simple example where \textsc{NReset} allows a second-law-violating cycle to be constructed: $ \langle NReset ; RPistIn ;  PartOut ;   RPistOut , \ (\tfrac{1}{2}, \ F, \ F, \ 1) \rangle \Rightarrow  (\tfrac{1}{2}, \ F, \ F, \ 2) $.

We can therefore state the following theorem:
\begin{theorem}
The operation \textsc{NReset} cannot be constructed out of the basic operations (PartIn, PartOut, LPistIn, LPistOut, RPistIn, RPistOut); whereas the Ladyman \textsc{Shift} operation can be so constructed. Furthermore, the addition of \textsc{NReset} to this basic set gives a set of operations that do not satisfy the second law of thermodynamics in its Kelvin form.
\end{theorem}

\section{Landauer Erasure}

We have shown that the basic set of operations (PartIn, PartOut, LPistIn, LPistOut, RPistIn, RPistOut) under the composition rules of \textsc{demonic} will always satisfy the second law. We now turn to the question that motivated the introduction of the language in the first place, and show that Landauer Erasure is also a necessary property of the system.

Key to understanding where this comes from is the invariant statement \eqref{inc}. Considering the two entropy quantities, we can describe them in terms of the different branches of the probabilistic computation. $\langle H(X) \rangle$ is the average entropy within a single branch of a computation, averaged across all branches, and $H(\langle X \rangle ) $ is the entropy of the distribution over all branches of the computation.

A simple way of intutitively distinguishing these quantities is to consider the case where $X=\tfrac{1}{2}$ and there is a partition, $A=T$. After a measurement of the particle position there will be two computational branches, in one of which $X=0$ and the other $X=1$. Before the measurement $\langle H(X) \rangle = H(\langle X \rangle )  = k \ln 2$. After the measurement $ \langle H(X) \rangle$ is now zero; however, $H(\langle X \rangle ) $ remains $k \ln 2$. If a further operation is performed that sends the $X=0$ state to $X=1$, then both entropy quantities will now be zero.

We have seen previously one example of a measurement operation within this system: applying a piston to a box with a partition in it. Whether the piston can in fact be inserted is a measurement of the position. If the position was uncertain beforehand, then the computation branches, with two possibilities for the two definite outcomes. This is measurement within this system: the probability distribution for the particle position beforehand then becomes the probability distribution over branches, in each of which the particle has a definite position, $X=0,1$. We can therefore define the action of measurement on the entropic quantities:  
\begin{definition} A \textit{measurement operation} has the effect of sending $\langle H(X) \rangle$ to $ 0$. It leaves $H(\langle X \rangle ) $ unchanged.
\label{mes}\end{definition}

One operation that can be performed once there has been a measurement is a reset operation: returning the system to a particular known state. This too can be defined entropically:
\begin{definition} A \textit{reset operation} acts on a post-measurement state, where $\langle H(X) \rangle = 0$, and has the effect of taking $H(\langle X \rangle )$ to $0$. It leaves $\langle H(X) \rangle $ unchanged.
\label{res}\end{definition}

Now we come to erasure. Erasure is the operation of returning the system to the definite known $X$ state $X=1$, from any starting state: $(X_0 , \ A_0, \ I_0, \ w_0) \longrightarrow (X=1 , \ A_f, \ I_f, \ w_f)$. There may be multiple branches to this computation, but in all of them $X=1$ in the final state. We therefore have the following definition: 

\begin{definition} An \textit{erasure operation} has the action of sending $\langle H(X) \rangle$ to $0$ and $H(\langle X \rangle )$ to $0$.
\label{era}\end{definition} 

From  Definitions \ref{mes}, \ref{res}, \ref{era}, we can conclude
\begin{lemma} An erasure operation is a measure and then reset operation. \label{lem1}\end{lemma}

We can now define minimum costs for these operations acting on certain states. The first interesting observation is that we recover a notion that was previously used to defeat the Maxwell Demon, but which fell out of favour once Landauer Erasure was postulated: that of a \textit{measurement cost} \cite{szilard}. Consider a single particle with $X=\tfrac{1}{2}$, i.e. its state is maximally unknown. The invariant statement tells us that any measurement operation must incur a work cost of $\tfrac{1}{2} k T \ln 2$ to reduce $\tfrac{1}{2}(\langle H(X) \rangle + H(\langle X \rangle )) $ from $k\ln 2 $ to $\tfrac{1}{2}k\ln2$. By Definition \ref{mes}, and using the computational invariant, we therefore have the following lemma:
\begin{lemma} A measurement operation of a single unknown bit of information takes $\langle H(X) \rangle = k\ln 2$ to $0$ but leaves $H(\langle X \rangle ) = k \ln 2$ unchanged. This incurs a \textbf{measurement work cost} of at least $\tfrac{1}{2}kT\ln2$.\label{lem2}
\end{lemma}

Defining a minimum erasure cost is slightly more complicated, in part owing to a lack of precision in the usual debate. Let us consider the measurement of a single bit as given above. Suppose we now wish to erase the result -- that is, we wish to reset the system in the known state $X=1$. We already have $\langle H(X) \rangle =0$ after the measurement; now we have also to take $H(\langle X \rangle ) = k \ln 2$ to $0$. By Definition \ref{res} and the computational invariant, we can conclude:
\begin{lemma} A reset operation for a single bit, acting on a measured bit that was unknown prior to the measurement, takes $H(\langle X \rangle ) = k \ln 2$ to 0  but leaves $\langle H(X) \rangle = 0$  unchanged. This incurs a \textbf{reset work cost} of at least $\tfrac{1}{2}kT\ln2$.\label{lem3}
\end{lemma}

We now consider a full erasure operation within single-particle thermodynamics: a measure--reset operation acting on an unknown bit ($X=\tfrac{1}{2}$). The bit value is measured, then the system reset into a known state. Using Definition \ref{era} and Lemmas \ref{lem1}, \ref{lem2}, and \ref{lem3}, we can state the main theorem:
\begin{theorem} An erasure operation for a single unknown bit of information takes both $\langle H(X) \rangle = k\ln 2$ and $H(\langle X \rangle ) = k \ln 2 $ to $ 0$. This incurs an \textbf{erasure work cost} of at least $kT\ln 2$.
\end{theorem}

Any erasure operation (measurement-and-reset) in the system described by \textsc{demonic} and the six basic operations will require an input of work into the system of at least $kT\ln2$: Landauer Erasure is a necessary part of this formalisation of single-particle thermodynamics.

\section{Conclusions}


We have introduced and formally specified a language for reasoning about single-particle equilibrium thermodynamics, \textsc{demonic}. The set of basic allowed operations, \textbf{Op}, is been described within this language, and we show that all programs formed by closing these basic operations under sequential composition and conditionals 
satisfy the probabilistic invariant statement $\langle w \rangle k \ln 2 - \tfrac{1}{2} \left( \langle H(X) \rangle + H(\langle X \rangle ) \right) \le 0 $. 

One consequence of this invariant statement is that any combination of such basic operations using the rules of the language necessarily forms a transition that satisfies the Kelvin statement of the second law of thermodynamics. The language and the  set \textbf{Op} can therefore be used to reason about thermodynamic properties without concern that a cycle violating the second law may inadvertently be constructed.


One unexpected consequence of the invariant statement is that \textit{measurement} can have an associated work cost. In particular, measuring an unknown bit costs at least $\tfrac{1}{2}kT\ln2$. A subsequent reset operation to the known state $X=1$ incurs a reset work cost of the same amount, $\tfrac{1}{2}kT\ln2$. Therefore, an erasure operation defined as measure-then-reset for a single unknown bit necessarily incurs a work cost of at least $kT\ln2$: Landauer erasure is a necessary consequence of the basic rules of the system, but only when combined with a measurement cost. The similarities with \cite{leffrex2} are suggestive.

By formalising and quantifying arguments that were previously largely linguistic, the programming approach of this paper enables clear conclusions to be drawn about previously debatable propositions. An important role is played by the probabilistic nature of some of some statements in the language, including some basic operations. By `keeping track' of all possibilities in the system as branches in a nondeterministic computation, this formalism enables us to distinguish two different entropic quantities: $\langle H(X) \rangle$ and $H(\langle X \rangle)$. The former is an average of the entropy of probability distributions for particle position within each branch/world; the latter is an entropy of the probability distribution across branches/worlds. Neither quantity alone satisfies an invariant proposition: both are required to generate the invariant statement, which can be viewed as the entropic/Clausius statement of the second law for this system.

It is important to note that we do not claim here either the second law or Landauer Erasure as theorems of the physical universe. What is presented here is a formal reasoning system that we claim is relevantly similar to toy model systems that are commonly used in arguments around the foundations of thermodynamics. The applicability of this \textit{model} to the actual world is a matter of ongoing debate -- especially around issues of statistical and/or quantum fluctuations -- and requires further research. 

There are many interesting questions that arise from this work. One is the relationship between the new entropic invariant $H(\langle X \rangle) + \langle H(X) \rangle$ and the Holevo quantity $H(\langle X \rangle) - \langle H(X) \rangle$ for a measurement \cite{holevo}, especially in light of the measurement work cost for this system. Another question is the exact relationship between the invariant statement and the second law in its Kelvin formulation. In this paper we have shown that the system satisfies both, and that the addition of an invariant-violating operation (\textsc{NReset}) also violates the second law. We have not, however, shown that there is an equivalence between the two constraints. There is also the extension of this language to larger systems: to more complicated probability distributions, and to systems with more than one particle and more than one box. The language and semantics have been specifically constructed with the aim of simplifying this extension. Finally, the exact relationship between the entropic quantities and probability distributions given here, and quantum states and entropies for single and multiple systems, will be a line of important further work.

The results presented here demonstrate the power of Computer Science methods for reasoning about processes in thermodynamics -- and have, potentially, a much broader range of applications in physics.\\

\noindent \textbf{Acknowledgements} D.C.H would like to thank Susan Stepney and Vaia Patta for interesting discussions and thoughtful comments on the text.

\nocite{*}
\bibliographystyle{eptcs}
\bibliography{dosbib}

\begin{thebibliography}{10}
\providecommand{\bibitemdeclare}[2]{}
\providecommand{\surnamestart}{}
\providecommand{\surnameend}{}
\providecommand{\urlprefix}{Available at }
\providecommand{\url}[1]{\texttt{#1}}
\providecommand{\href}[2]{\texttt{#2}}
\providecommand{\urlalt}[2]{\href{#1}{#2}}
\providecommand{\doi}[1]{doi:\urlalt{http://dx.doi.org/#1}{#1}}
\providecommand{\bibinfo}[2]{#2}

\bibitemdeclare{incollection}{giry1982categorical}
\bibitem{giry1982categorical}
\bibinfo{author}{Michele \surnamestart Giry\surnameend} (\bibinfo{year}{1982}):
  \emph{\bibinfo{title}{A categorical approach to probability theory}}.
\newblock In: {\sl \bibinfo{booktitle}{Categorical aspects of topology and
  analysis}}, \bibinfo{publisher}{Springer}, pp. \bibinfo{pages}{68--85},
  \doi{10.1007/BFb0092872}.

\bibitemdeclare{book}{hennessy1990semantics}
\bibitem{hennessy1990semantics}
\bibinfo{author}{M~\surnamestart Hennessy\surnameend} (\bibinfo{year}{1990}):
  \emph{\bibinfo{title}{The semantics of programming languages: an elementary
  introduction using structural operational semantics}}.
\newblock \bibinfo{publisher}{John Wiley \& Sons}.

\bibitemdeclare{article}{hoare}
\bibitem{hoare}
\bibinfo{author}{C.~A.~R.~\surnamestart Hoare\surnameend} (\bibinfo{year}{1969}):
  \emph{\bibinfo{title}{An axiomatic basis for computer programming}}.
\newblock {\sl \bibinfo{journal}{Comm. {ACM}}} \bibinfo{volume}{12}, pp.
  \bibinfo{pages}{576--580}, \doi{10.1145/363235.363259}.

\bibitemdeclare{article}{holevo}
\bibitem{holevo}
\bibinfo{author}{Alexander~S. \surnamestart Holevo\surnameend}
  (\bibinfo{year}{1973}): \emph{\bibinfo{title}{Bounds for the quantity of
  information transmitted by a quantum communication channel}}.
\newblock {\sl \bibinfo{journal}{Problems of Information Transmission}}
  \bibinfo{volume}{9}, pp. \bibinfo{pages}{177Ð--183}.

\bibitemdeclare{article}{ladyman08}
\bibitem{ladyman08}
\bibinfo{author}{J.~\surnamestart Ladyman\surnameend},
  \bibinfo{author}{S.~\surnamestart Presnell\surnameend} \&
  \bibinfo{author}{A.~J. \surnamestart Short\surnameend}
  (\bibinfo{year}{2008}): \emph{\bibinfo{title}{The use of the
  information-theoretic entropy in thermodynamics}}.
\newblock {\sl \bibinfo{journal}{Stud. Hist. Phil. Mod. Phys.}}
  \bibinfo{volume}{39}, pp. \bibinfo{pages}{315--324},
  \doi{10.1016/j.shpsb.2007.11.004}.

\bibitemdeclare{article}{ladyman07}
\bibitem{ladyman07}
\bibinfo{author}{J.~\surnamestart Ladyman\surnameend},
  \bibinfo{author}{S.~\surnamestart Presnell\surnameend},
  \bibinfo{author}{A.~J. \surnamestart Short\surnameend} \&
  \bibinfo{author}{B.~\surnamestart Groisman\surnameend}
  (\bibinfo{year}{2007}): \emph{\bibinfo{title}{The connection between logical
  and thermodynamic irreversibility}}.
\newblock {\sl \bibinfo{journal}{Stud. Hist. Phil. Mod. Phys.}}
  \bibinfo{volume}{38}, pp. \bibinfo{pages}{58--79},
  \doi{10.1016/j.shpsb.2006.03.007}.

\bibitemdeclare{article}{ladyman13}
\bibitem{ladyman13}
\bibinfo{author}{J.~\surnamestart Ladyman\surnameend} \&
  \bibinfo{author}{K.~\surnamestart Robertson\surnameend}
  (\bibinfo{year}{2013}): \emph{\bibinfo{title}{Landauer defended: reply to
  {N}orton}}.
\newblock {\sl \bibinfo{journal}{Stud. Hist. Phil. Mod. Phys.}}
  \bibinfo{volume}{44}, pp. \bibinfo{pages}{263--271},
  \doi{10.1016/j.shpsb.2013.02.005}.

\bibitemdeclare{article}{landauer}
\bibitem{landauer}
\bibinfo{author}{R.~\surnamestart Landauer\surnameend} (\bibinfo{year}{2061}):
  \emph{\bibinfo{title}{Irreversibility and heat generation in the computing
  process}}.
\newblock {\sl \bibinfo{journal}{{IBM} Journ. of R\& D}} \bibinfo{volume}{5},
  pp. \bibinfo{pages}{183--191}, \doi{10.1147/rd.53.0183}.

\bibitemdeclare{article}{leffrex2}
\bibitem{leffrex2}
\bibinfo{author}{H.~\surnamestart Leff\surnameend} \&
  \bibinfo{author}{A.~\surnamestart Rex\surnameend} (\bibinfo{year}{1994}):
  \emph{\bibinfo{title}{Entropy of measurement and erasure: {S}zilard's
  membrane revisited}}.
\newblock {\sl \bibinfo{journal}{Am. Journ. Phys.}} \bibinfo{volume}{63}, pp.
  \bibinfo{pages}{994--1000}, \doi{10.1119/1.17749}.

\bibitemdeclare{book}{leffrex}
\bibitem{leffrex}
\bibinfo{author}{H.~\surnamestart Leff\surnameend} \&
  \bibinfo{author}{A.~\surnamestart Rex\surnameend} (\bibinfo{year}{2003}):
  \emph{\bibinfo{title}{Maxwell's Demon 2}}.
\newblock \bibinfo{publisher}{IOP}.

\bibitemdeclare{inproceedings}{toffoli}
\bibitem{toffoli}
\bibinfo{author}{L.B. \surnamestart Levitin\surnameend} \&
  \bibinfo{author}{T.~\surnamestart Toffoli\surnameend} (\bibinfo{year}{2006}):
  \emph{\bibinfo{title}{Thermodynamic Cost of Reversible Computing}}.
\newblock In: {\sl \bibinfo{booktitle}{Information Theory, 2006 IEEE
  International Symposium on}}, pp. \bibinfo{pages}{2082--2084},
  \doi{10.1109/ISIT.2006.261917}.

\bibitemdeclare{article}{lloyd}
\bibitem{lloyd}
\bibinfo{author}{S.~\surnamestart Lloyd\surnameend} (\bibinfo{year}{1997}):
  \emph{\bibinfo{title}{Quantum mechanical {M}axwell's demon}}.
\newblock {\sl \bibinfo{journal}{Phys. Rev. A}} \bibinfo{volume}{56}, pp.
  \bibinfo{pages}{3374--3382}, \doi{10.1103/PhysRevA.56.3374}.

\bibitemdeclare{article}{maroney05}
\bibitem{maroney05}
\bibinfo{author}{O.~\surnamestart Maroney\surnameend} (\bibinfo{year}{2005}):
  \emph{\bibinfo{title}{The (absense of a) relationship between thermodynamic
  and logical irreversibility}}.
\newblock {\sl \bibinfo{journal}{Stud. Hist. Phil. Mod. Phys.}}
  \bibinfo{volume}{36}, pp. \bibinfo{pages}{355Ð--374},
  \doi{10.1016/j.shpsb.2004.11.006}.

\bibitemdeclare{book}{maxwell}
\bibitem{maxwell}
\bibinfo{author}{J.~C. \surnamestart Maxwell\surnameend}
  (\bibinfo{year}{1871}): \emph{\bibinfo{title}{Theory of Heat}}.
\newblock \bibinfo{publisher}{Longman, Green, and co. London}.

\bibitemdeclare{article}{norton11}
\bibitem{norton11}
\bibinfo{author}{J.~\surnamestart Norton\surnameend} (\bibinfo{year}{2011}):
  \emph{\bibinfo{title}{Waiting for {L}andauer}}.
\newblock {\sl \bibinfo{journal}{Stud. Hist. Phil. Mod. Phys.}}
  \bibinfo{volume}{42}, pp. \bibinfo{pages}{184--198},
  \doi{10.1016/j.shpsb.2011.05.002}.

\bibitemdeclare{book}{plotkin1981structural}
\bibitem{plotkin1981structural}
\bibinfo{author}{G.~D. \surnamestart Plotkin\surnameend}
  (\bibinfo{year}{1981}): \emph{\bibinfo{title}{A structural approach to
  operational semantics}}.
\newblock \bibinfo{publisher}{DAIMI Aarhus University}.

\bibitemdeclare{article}{szilard}
\bibitem{szilard}
\bibinfo{author}{L.~\surnamestart Szilard\surnameend} (\bibinfo{year}{1925}):
  \emph{\bibinfo{title}{On the decrease of entropy in a themodynamic system by
  the intervention of intelligent beings}}.
\newblock {\sl \bibinfo{journal}{Z. fur Physik}} \bibinfo{volume}{32}, pp.
  \bibinfo{pages}{753--788}, \doi{10.1007/BF01331713}.

\bibitemdeclare{article}{vlatko}
\bibitem{vlatko}
\bibinfo{author}{V.~\surnamestart Vedral\surnameend} (\bibinfo{year}{2000}):
  \emph{\bibinfo{title}{Landauer's erasure, error correction, and
  entanglement}}.
\newblock {\sl \bibinfo{journal}{Proc. R. Soc. Lond. A}} \bibinfo{volume}{456},
  pp. \bibinfo{pages}{969--984}, \doi{10.1098/rspa.2000.0545}.

\end{thebibliography}

\newpage

\section*{Appendix: Proof of the Computational Invariant}

We prove that the statement \eqref{inc},
$$ \mathrm{P}: \ \ \langle wk\ln 2 \rangle - \tfrac{1}{2} \left( \langle H(X) \rangle + H(\langle X \rangle ) \right) \le 0 $$

\noindent is a computational invariant of operations $Op_i \in \{\text{PartIn}, \text{PartOut}, \text{LPistIn}, \text{LPistOut}, \text{RPistIn}, \text{RPistOut}\}$ and their composition under the semantics of \textsc{demonic}. \\

Using the standard inference rules of Hoare logic we show for each operation that P $\{ Op_i \}$ P. We will use the shorthand $ 2\Sigma(X) = \langle H(X) \rangle + H(\langle X \rangle )$.\\

To begin, we note that P $\{ s.A=B \}$ P and P $\{ s.I=B \}$ P follow trivially from the assignment axiom $\mathrm{P}[E/x] \{x := E\} \mathrm{P} \ $ as P is independent of $I$ and $A$. We therefore only need to check the validity of assignments for $X$ and $w$.\\

\textbf{PartIn} $\ =_{def}\ (s.A := \mathtt{true})$\\

\noindent P $\{$PartIn$\}$ P is trivially true as there is no assignment that changes $X$ or $w$. \qed\\

%
%

\textbf{PartOut}:\\
\indent \hspace{2cm} $\langle s = (X_0 , A_0, I_0, w_0) \rangle$\\
\indent \hspace{2cm} if $(s.A = \mathtt{true})$ then \\
\indent \hspace{3cm}  $\langle s = (X_0 , \mathtt{true}, I_0, w_0) \rangle$\\
\indent \hspace{3cm} (if $(s.I = \mathtt{false})$ then\\
\indent \hspace{4cm}  $\langle s = (X_0 , \mathtt{true}, \mathtt{false}, w_0) \rangle$\\
\indent \hspace{4cm} $(s.X := \tfrac{1}{2}) \ $ and $ (s.A := \mathtt{false}) \ $ else\\
\indent \hspace{5cm}  $\langle s = (X_0 , \mathtt{true}, \mathtt{true}, w_0) \rangle$\\
\indent \hspace{5cm} $(s.A := \mathtt{false}) $ )\\
\indent \hspace{6cm}  $\langle s = (X_0 , \mathtt{false}, I_0, w_0) \rangle$\\
\indent \hspace{6cm} else $\mathtt{skip}$\\

\noindent The one assignment that changes a variable in P is $s.X := \tfrac{1}{2}$, and it acts when the state is $(X_0 , \mathtt{true}, \mathtt{false},$ $w_0)$. The assignment axiom gives
\begin{equation} \langle wk\ln 2 \rangle - \Sigma(\tfrac{1}{2}) \le 0 \ \{ s.X:=\tfrac{1}{2} \} \ \langle wk\ln 2 \rangle - \Sigma(X) \le 0 \label{partout1} \end{equation}

\noindent By definition of $\Sigma$, within the system given
\begin{equation} \forall X_0 \ \ \ \Sigma(\tfrac{1}{2}) \ge \Sigma(X_0)\end{equation}

\noindent Therefore
\begin{equation} \langle wk\ln 2 \rangle - \Sigma(X_0) \le 0 \ \ \Longrightarrow \ \ \langle wk\ln 2 \rangle - \Sigma(\tfrac{1}{2}) \le 0\label{partout2}
\end{equation}

\noindent From \eqref{partout2} and \eqref{partout1} we conclude
\begin{equation} \langle wk\ln 2 \rangle - \Sigma(X_0) \le 0 \ \{ s.X:=\tfrac{1}{2} \} \ \langle wk\ln 2 \rangle - \Sigma(X) \le 0  \end{equation}

\noindent Therefore P $\{$PartOut$\}$ P as required. \qed

\textbf{LPistOut}:\\
\indent \hspace{1cm} $\langle s = (X_0 , A_0, I_0, w_0) \rangle$\\
\indent \hspace{1cm} if $(s.I = \mathtt{false})$ or $\neg (s.X = 0)$ then \\
\indent \hspace{2cm} $\mathtt{skip}$ else\\
\indent \hspace{3cm} $\langle s = (0 , A_0, T, w_0) \rangle$\\
\indent \hspace{3cm} ( if $(s.A = \mathtt{true}) \ $ then\\
\indent \hspace{4cm} $\langle s = (0 , T, T, w_0) \rangle$\\
\indent \hspace{4cm} $(s.I := \mathtt{false}) \ $ else\\
\indent \hspace{5cm} $\langle s = (0 , F, T, w_0) \rangle$\\
\indent \hspace{5cm} $(s.I := \mathtt{false}) \ $ and $(s.X := \frac{1}{2}) \ $ and $(s.w := w+1)$ )\\

\noindent The assignment given in the final line is the only one we need to verify. Two uses of the assignment axiom gives us
\begin{equation} \langle (w+1)k\ln 2 \rangle - \Sigma(\tfrac{1}{2}) \le 0 \ \{ s.X:=\tfrac{1}{2} \ ; \ s.w := w+1\} \ \langle wk\ln 2 \rangle - \Sigma(X) \le 0 \label{pistout1} \end{equation}

\noindent The state the assignment under consideration is acting on is $(0 , F, T, w_0)$. In that case, the invariant beforehand is $ \langle w_0k\ln 2 \rangle \le 0$, from which (recalling $\Sigma(\tfrac{1}{2})=k\ln 2$) we can infer

\begin{equation} \langle w_0k\ln 2 \rangle \le 0 \ \Longrightarrow \ \langle (w_0+1)k\ln 2 \rangle - \Sigma(\tfrac{1}{2}) \le 0
\label{pistout2}\end{equation}

\noindent From \eqref{pistout2} and \eqref{pistout1} we can conclude
\begin{equation} \langle w_0k\ln 2 \rangle - \Sigma(0) \le 0 \ \{ s.X:=\tfrac{1}{2} \ ; \ s.w := w+1 \} \ \langle wk\ln 2 \rangle - \Sigma(X) \le 0  \end{equation}

\noindent Therefore P $\{$PistOut$\}$ P as required. \qed\\

\textbf{RPistOut}:

The proof proceeds as for LPistOut by symmetry. \qed \\

\textbf{LPistIn}:\\
\indent \hspace{1cm} $\langle s = (X_0 , A_0, I_0, w_0) \rangle$\\
\indent \hspace{1cm} if  $(s.X = 1)\ $ then \\
\indent \hspace{1.5cm} $\langle s = (1 , A_0, I_0, w_0) \rangle$\\
\indent \hspace{1.5cm} $(s.w := w-1)\ $ else\\
\indent \hspace{2cm} ( if $(s.X = 0)\ $ then\\
\indent \hspace{2.5cm} $\langle s = (0 , A_0, I_0, w_0) \rangle$\\
\indent \hspace{2.5cm} $(s.I := \mathtt{true}) \ $ else\\
\indent \hspace{3cm} $\langle s = (\tfrac{1}{2} , A_0, I_0, w_0) \rangle$\\
\indent \hspace{3cm} ( if $ (s.A = \mathtt{false})\ $ then\\
\indent \hspace{3.5cm} $\langle s = (\tfrac{1}{2} , F, I_0, w_0) \rangle$\\
\indent \hspace{3.5cm} $(s.X := 1)\ $ and $ (s.w := w-1) \ $ and $(s.I := \mathtt{true})\ $ else\\
\indent \hspace{4cm} $\langle s = (\tfrac{1}{2} , T, I_0, w_0) \rangle$\\
\indent \hspace{4cm} $[(s.X := 0) \ \mathrm{and} \ (s.I := \mathtt{true})] \oplus [(s.X := 1) \ \mathrm{and} \ (s.w := w-1)]$))\\

\noindent The final two statements are the ones we need to verify.\\

\noindent \textit{Statement 1}: $(s.X := 1)\ $ and $ (s.w := w-1) \ $ in the state $\langle s = (\tfrac{1}{2} , F, I_0, w_0) \rangle$.\\

\noindent Two uses of the assignment axiom give us
\begin{equation} \langle (w-1)k\ln 2 \rangle - \Sigma(1) \le 0 \ \{ s.X:=1 \ ; \ s.w := w-1\} \ \langle wk\ln 2 \rangle - \Sigma(X) \le 0 \label{pistin11} \end{equation}

\noindent In the state $(\tfrac{1}{2} , F, I_0, w_0)$ we have $ \langle w_0k\ln 2 \rangle - \Sigma(\tfrac{1}{2}) \le 0$. Recalling $\Sigma(\tfrac{1}{2})=k\ln 2$ and $\Sigma(1)=\Sigma(0)=0$, we have
\begin{equation} \langle w_0k\ln 2 \rangle - \Sigma(\tfrac{1}{2}) \le 0 \ \Longrightarrow \ \langle (w_0-1)k\ln 2 \rangle - \Sigma(1) \le 0
\label{pistin12}\end{equation}

\noindent From \eqref{pistin12} and \eqref{pistin11} we can conclude
\begin{equation} \langle w_0k\ln 2 \rangle - \Sigma(\tfrac{1}{2}) \le 0 \ \{ s.X:=1 \ ; \ s.w := w-1 \} \ \langle wk\ln 2 \rangle - \Sigma(X) \le 0  \label{pistin1}\end{equation}

\noindent \textit{Statement 2}: $[(s.X := 0) ] \oplus [(s.X := 1) \ \mathrm{and} \ (s.w := w-1)]$ in the state $ (\tfrac{1}{2} , T, I_0, w_0) $.\\

We use a different proof method as the transition is probabilistic: we show directly that the value of the invariant quantity does not change in this transition. The transition gives the 50-50 probabilistic mixture afterwards
\begin{equation} (X=\tfrac{1}{2},\ \Sigma(X) = k\ln 2, \ w=w_0)  \hspace{.5cm} \Longrightarrow_\frac{1}{2} \hspace{.5cm} 
\left[ \begin{array}{l} (X=1, \ w=w_0-1) \\ 
 \  (X=0, \ w=w_0) \end{array}\right]
 \end{equation}

\noindent Note we have not yet given the value for $\Sigma(X)$ at the end of the transition, as this depends on both branches of the outcome. This quantity is
\begin{align} \Sigma(X_{out})  &= \tfrac{1}{2} \Big( \ \langle H(X_{out}) \rangle + H(\langle X_{out} \rangle ) \ \Big)\nonumber\\
&= \tfrac{1}{2} \Big( \tfrac{1}{2} ( H(0) + H(1) ) + H( \tfrac{1}{2} )  \ \Big)\nonumber\\
&= \tfrac{1}{2} k \ln 2 \end{align}

\noindent To compute the invariant afterwards, we therefore have
\begin{equation} \langle w_{out} \rangle k \ln 2 - \Sigma(X_{out}) = \langle w_{in} \rangle k \ln 2 -  \tfrac{1}{2}k\ln 2 - \Big( \Sigma(X_{in}) - \tfrac{1}{2}k\ln 2 \Big) = \langle w_{in} \rangle k \ln 2 - \Sigma(X_{in}) \end{equation}

\noindent It follows straightforwardly that
 \begin{equation} \langle w_0k\ln 2 \rangle - \Sigma(\tfrac{1}{2}) \le 0 \ \{ [(s.X := 0) ] \oplus [(s.X := 1) \ \mathrm{and} \ (s.w := w-1)] \} \ \langle wk\ln 2 \rangle - \Sigma(X) \le 0  \label{pistin2}\end{equation}
 
\noindent From \eqref{pistin1} and \eqref{pistin2} we can conclude P $\{$LPistIn$\}$ P as required. \qed\\

\textbf{RPistIn}:

The proof proceeds as for LPistIn by symmetry. \qed \\

\end{document}